\documentclass[pre,aps,amsmath,twocolumn,showpacs,superscriptaddress,
longbibliography]{revtex4-1}

\usepackage{amssymb}
\usepackage{graphicx}
\usepackage[pdftex,bookmarks,colorlinks,breaklinks]{hyperref}
\hypersetup{linkcolor=red,citecolor=blue,filecolor=dullmagenta,urlcolor=blue}
\usepackage{color}

\usepackage{soul}
\definecolor{indiagreen}{rgb}{0.07, 0.53, 0.03}

\usepackage{amsmath}
\usepackage{mathrsfs}
\usepackage{amsfonts}

\begin{document}

\title{Microwave graphs analogs for the voltage drop in three-terminal devices 
with orthogonal, unitary and symplectic symmetry}

\author{F. Casta\~neda-Ram\'irez}
\affiliation{Departamento de F\'isica, Universidad Aut\'onoma 
Metropolitana-Iztapalapa, Apartado Postal 55-534, 09340 Ciudad de M\'exico, 
Mexico}

\author{A.~M. Mart\'inez-Arg\"uello}
\email{blitzkriegheinkel@gmail.com}
\affiliation{Instituto de Ciencias F\'isicas, Universidad Nacional Aut\'onoma 
de M\'exico, Apartado Postal 48-3, 62210 Cuernavaca, Mor., Mexico}

\author{T. Hofmann}
\address{Fachbereich Physik der Philipps-Universit\"at Marburg, D-35032
Marburg, Germany}

\author{A. Rehemanjiang}
\affiliation{Fachbereich Physik der Philipps-Universit\"at Marburg, D-35032
Marburg, Germany}

\author{M. Mart\'{\i}nez-Mares}
\affiliation{Departamento de F\'isica, Universidad Aut\'onoma 
Metropolitana-Iztapalapa, Apartado Postal 55-534, 09340 Ciudad de M\'exico, 
Mexico}

\author{J.~A. M\'endez-Berm\'udez}
\affiliation{Instituto de F\'isica, Benem\'erita Universidad Aut\'onoma de Puebla, 
Apartado Postal J-48, 72570 Puebla, Pue., Mexico}

\author{U. Kuhl}
\affiliation{Universit\'{e} C\^{o}te d'Azur, CNRS, Institut de Physique de Nice (INPHYNI), 06108 Nice, France, EU}

\author{H.-J. St\"ockmann}
\affiliation{Fachbereich Physik der Philipps-Universit\"at Marburg, D-35032
Marburg, Germany}


\begin{abstract}

Transmission measurements through three-port microwave graphs are performed in
a symmetric setting, in analogy to three-terminal voltage drop devices with
orthogonal, unitary, and symplectic symmetry. The terminal
used as a probe is symmetrically located between two chaotic graphs, each graph
is connected to one port, the input and the output, respectively. The analysis 
of the experimental data exhibit the weak localization and antilocalization 
phenomena in a clear fashion. We find a good agreement with theoretical 
predictions, provided that the effect of dissipation and imperfect coupling 
to the ports are taken into account.

\end{abstract}


\maketitle


\section{Introduction}

The wave origin of quantum interference in many physical phenomena opens the
possibility that analogous classical wave systems may emulate quantum
devices~\cite{Fyodorov2005a}. For example, classical wave systems have been used as
auxiliary tools to understand transport properties of multiterminal quantum
systems. The dissipation that occurs in classical wave systems and the
imperfect coupling to the leads that feed the system have not been
disadvantageous, but interesting phenomena deserving to be studied
to analyze their effects in transport properties~\cite{Schanze2001,Schanze2005,Schafer2003,Kuhl2005}.

An opportunity to study many electrical or thermal conduction properties of 
solid state physics by classical wave systems has been opened once the problem 
of electrical or thermal conduction is reduced to a scattering problem. For
instance, Landauer formula states that the electrical conductance is
proportional to the transmission coefficient~\cite{Buttiker1988}; an equivalent
Landauer formula is valid for thermal conductance~\cite{Schwab2000}. Therefore, a
lot of research has been devoted to the study of conduction through two-terminal 
configurations from the theoretical and experimental points of view, using 
quantum
mechanics~\cite{Buttiker1984,Brouwer1997,Keller1996,Marcus1992,Chan1995} as well as
classical wave physics~\cite{Schanze2001,Schanze2005,Moises2005,Enrique2016}.

Multiterminal systems have also been studied 
in~\cite{Buttiker1986,Gopar1994,Arrachea2008,Texier2016,Foieri2009,DAmato1990,
Cattena2014,Song1998,Gao2005}. 
Among them three-terminal systems have been considered in which the voltage drop 
along the system is the observable of interest~\cite{Angel2018,Angel2019}. While 
two of the terminals are connected to fixed electronic reservoirs to feed the 
system, the third one is used as a probe that tunes the voltage drop. The 
voltage drop along a disordered wire was analyzed for the one-channel 
case~\cite{Godoy1992a,Godoy1992b}. It has been shown theoretically that the 
value of voltage drop lies between those on the two terminals; in fact, it 
consists of the average of the voltages in the terminals and a deviation term, 
which contains specific information of the system through the transmsission 
coefficients from the terminals to the probe. This \emph{voltage drop 
deviation} term, $f$, is an important quantity in the study of the voltage drop 
reducing the problem to a scattering problem. It has been shown that $f$ 
fluctuates from sample-to-sample and its statistical distribution has a 
completely different shape for metallic and insulating 
regimes~\cite{Godoy1992a,Godoy1992b}. The same quantity was also considered for 
a chaotic system, numerically simulated using Random Matrix Theory 
(RMT)~\cite{Angel2014}.

More recently, an experiment with microwave graphs has been performed for an
asymmetric configuration, where the probe is located on one side of a chaotic
graph~\cite{Angel2018}. Analytical and numerical procedures were carried out to 
compute $f$. Remarkably, the results show significant differences with respect 
to the corresponding disordered case. Here, we deepen in the understanding of 
the voltage drop by proposing an experiment with microwave graphs but locating 
the probe between two chaotic graphs, as shown in Fig.~\ref{fig:middle3T}. Our 
proposal is accompanied by analytical predictions.

The paper is organized as follows. In the next section we obtain the
analytical expressions for the voltage drop for a measurement at the middle
of two scattering devices. It is written in terms of the scattering matrices of 
the individual devices. In section~\ref{sec:Chaos} the analytical procedure, 
using RMT calculations for the scattering matrices, is developed for the 
statistical distribution of the deviation $f$ assuming chaotic scattering 
devices. The experimental realization with microwave graphs is explained in 
section~\ref{sec:exp}, where we compare experimental results with theoretical 
predictions. Finally, we present our conclusions in
section~\ref{sec:conclusions}.


\section{Voltage drop reduced to a scattering problem}


\subsection{Voltage drop at the middle of a symmetric device}

\begin{figure}
\begin{center}
\includegraphics[width=1.0\columnwidth]{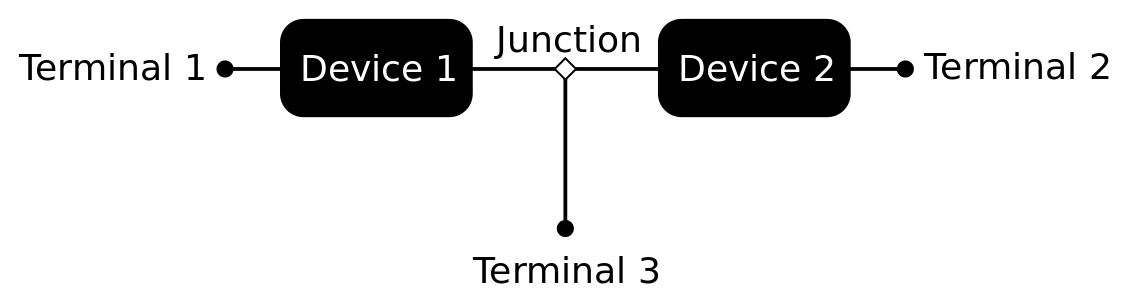}
\caption{Schematic representation of the three-terminal system for measuring the 
voltage drop through the third terminal located between of two chaotic devices.}
\label{fig:middle3T}
\end{center}
\end{figure}

To measure the voltage drop along a quantum wire in the simplest configuration, 
a three-terminal system is needed in which one of the ports is used as a probe 
by tuning its voltage to zero current. A schematic view of the system where the 
probe is located between two quantum devices is shown in 
Fig.~\ref{fig:middle3T}. We are interested in the one-channel situation of the 
leads connecting the system, for  which the zero current in the probe implies 
that~\cite{Buttiker1988}
\begin{equation}
\mu_3 = \frac{1}{2}\left( \mu_1 + \mu_2\right) +
\frac{1}{2} \left( \mu_1 - \mu_2 \right)\, f,
\label{eq:mu3}
\end{equation}
where $\mu_i$ is the chemical potential of the $i$th electronic reservoir and
\begin{equation}
f = \frac{T_{31} - T_{32}}{T_{31} + T_{32}},
\label{eq:f}
\end{equation}
with $T_{ij}$ the transmission coefficient from Terminal $j$ to Terminal
$i$. Since $f$ gives a measure of the deviation of the voltage drop from the
average value of potentials $\mu_1$ and $\mu_2$, which is obtained in the ideal
case where the two devices are straight waveguides, we will refer to it as
the \emph{voltage drop deviation}; it takes values in the interval [-1, 1].

\subsection{Voltage drop deviation $f$ in terms of scattering elements}

Equation (\ref{eq:f}) is equivalent to the conductance but for three-terminal
devices. Analogously to the conductance the problem of the voltage drop in
electronic devices is reduced to a scattering problem through the quantity $f$.
It is clear that $f$ contains all the information about the system by means of the
transmission coefficients $T_{ij}$. Let us assume that the scattering properties
of each device are known through their scattering matrices, $S_1$ for Device 1
and $S_2$ for Device 2. The general structure of these matrices depends on the 
symmetry properties of the problem, namely
\begin{equation}
S_j =
\left(
\begin{array}{cc}
r_{j}J & t'_{j}D_j \\
t_{j}D_j^R & r'_{j}J
\end{array}
\right), \quad j=1,\,2,
\end{equation}
where $r_j$ ($r'_j$) and $t_j$ ($t'_j$) are the reflection and transmission 
amplitudes for incidence from the left (right) of the Device $j$. 
Unitarity is the only requirement on $S_{j}$ in absence of any 
symmetry; this case is known as the Unitary symmetry and is labeled by $\beta=2$
in Dyson's scheme. In addition, in the presence of time reversal
invariance $S_{j}$ is unitary and symmetric, which corresponds to the Orthogonal 
symmetry and is labeled by $\beta=1$; for both cases $D_j=D_j^R=J=1$. 
Furthermore, in the presence of time-reversal invariance but no spin-rotation 
symmetry, the symmetry of the system is the Symplectic one, labeled by 
$\beta=4$, and $S_j$ becomes a $4\times 4$ self-dual matrix~\cite{RMP69} in 
which case $J$ is the $2\times 2$ identity matrix, $J=\openone$, and $D_j^R$ is 
the dual matrix of the $2\times 2$ matrix $D_j$, defined by~\cite{Karol}
\begin{equation}
\label{eq:dual}
D_j^R = -ZD_j^TZ,
\end{equation}
with $D_j^T$ the transposed matrix of $D_j$ and
\begin{equation}
Z =
\left(
\begin{array}{cc}
0 & -1 \\
1 & 0
\end{array}
\right).
\end{equation}

The dependence of $T_{ij}$ on $S_1$ and $S_2$  can be obtained explicitly.
The scattering matrix of the three-terminal symmetric configuration is given by~\cite{Angel2014}
\begin{equation}
\mathbf{S} = S_{PP}+S_{PQ}(S_0-S_{QQ})^{-1}S_{QP},
\label{S}
\end{equation}
where $S_0$ is the scattering matrix of the junction that accounts for the
coupling to the probe. In fact, $S_0$ can be obtained from the
experiment~\cite{Angel2018,Abdu2018}, it reads as
\begin{equation}
\label{eq:S0}
S_{0}= \frac{1}{3} \left(
\begin{array}{ccc}
-J & 2J & 2J \\
2J & -J & 2J \\
2J & 2J & -J
\end{array}
\right) .
\end{equation}

In Eq.~(\ref{S}) $S_{PP}$ represents the reflections to the terminals, $S_{QP}$ 
and $S_{PQ}$ represent the transmissions from the terminals to the inner part of 
the system and from the inner part to the terminals; $S_{QQ}$ represents the 
internal reflections and $(S_0-S_{QQ})^{-1}$ accounts for the multiple 
scattering between the Devices and the junction. Explicitly, 
they are
\begin{equation}
\begin{array}{cc}
\vspace{0.5cm}
S_{PP} = \left(
\begin{array}{ccc}
r_{1}J & 0 & 0 \\
0 & r'_{2}J & 0 \\
0 & 0 & 0
\end{array}
\right), & S_{PQ} = \left(
\begin{array}{ccc}
t'_{1}D_1 & 0 & 0 \\
0 & t_2D^R_2 & 0 \\
0 & 0 & J
\end{array}
\right), \\
S_{QP} = \left(
\begin{array}{ccc}
t_{1}D^R_1 & 0 & 0 \\
0 & t'_{2}D_2 & 0 \\
0 & 0 & J
\end{array}
\right), & S_{QQ} = \left(
\begin{array}{ccc}
r'_{1}J & 0 & 0 \\
0 & r_{2}J & 0 \\
0 & 0 & 0
\end{array}
\right).
\end{array}
\label{eq:Ss}
\end{equation} 
By substituting Eqs.~(\ref{eq:S0}) and (\ref{eq:Ss}) into Eq.~(\ref{S}), the 
scattering matrix elements $\mathbf{S}_{31}$ and $\mathbf{S}_{32}$ are 
obtained, from which $T_{ij}=|\mathbf{S}_{ij}|^2$ for $\beta=1,\, 2$, while 
$T_{ij}=\frac{1}{2}\mathrm{tr}(\mathbf{S}_{ij}\mathbf{S}^{\dagger}_{ij})$ for 
$\beta=4$, and therefore Eq.~(\ref{eq:f}) reduces to
\begin{equation}
\label{eq:f3}
f = \frac{|t_1|^2|1+r_2|^2-|t'_2|^2|1+r'_1|^2}
{|t_1|^2|1+r_2|^2+|t'_2|^2|1+r'_1|^2}.
\end{equation}
Note that $S_{31}$, $S_{32}$, and $f$ depend only on the elements of the 
individual scattering matrices that describe the devices.


\section{Chaotic scattering for the voltage drop deviation}
\label{sec:Chaos}

Of particular interest are the transport properties through chaotic devices,
quantum or classical. The disordered three-terminal system was previously
considered in Ref.~\cite{Godoy1992b}. In any case the scattering quantities, like
the transmission coefficients through each device, fluctuate with respect to
a tuning parameter, like the energy of the incident particles in the quantum
case or the frequency in the classical wave situation, or from sample to
sample. Therefore, it is the distribution of $f$ what becomes more important
rather than a particular value of $f$.

For a chaotic cavity, the statistical fluctuations of the scattering matrix are
described by RMT. There, $S_j$ is uniformly distributed according to the 
invariant measure $\mathrm{d}\mu_{\beta}(S_j)$ that defines the circular 
ensemble for the symmetry class $\beta$: the Circular Orthogonal Ensemble for 
$\beta=1$, the Circular Unitary Ensemble for $\beta=2$, and the Circular
Symplectic Ensemble for $\beta=4$. Hence, the statistical distribution of $f$
can be calculated from the definition
\begin{equation}
P_\beta(f) = \int
\delta\left(f-\frac{T_{31}-T_{32}}{T_{31}+T_{32}}\right)
\mathrm{d}\mu_{\beta} (S_1)\, \mathrm{d}\mu_{\beta}(S_2) ,
\label{eq:Pf1}
\end{equation}
where $\delta$ is the Dirac delta function.


\subsection{Statistical distribution of the voltage drop deviation $f$}

A useful parameterization for scattering matrices is the polar form
\begin{equation}
S_j =
\left[
\begin{array}{cc}
-\sqrt{1-\tau_j}\textnormal{e}^{\textnormal{i}(\phi_j+\phi'_j)}J &
\sqrt{\tau_j}\textnormal{e}^{\textnormal{i}(\phi_j+\psi'_j)}D_j \\
\sqrt{\tau_j}\textnormal{e}^{\textnormal{i}(\psi_j+\phi'_j)}D_j^R  &
\sqrt{1-\tau_j}\textnormal{e}^{\textnormal{i}(\psi_j+\psi'_j)}J
\end{array}
\right],
\label{SjP}
\end{equation}
where $0\leq\tau_j\leq1$ and $\phi_j$, $\phi'_j$, $\psi_j$, $\psi'_j$ lie in
the interval $[0,\, 2\pi]$; $\phi_j=\phi'_j$ and $\psi_j=\psi'_j$ for
$\beta=1,\,4$. Also, for $\beta=4$, $D_j$ can be parameterized as 
\begin{equation}
D_j=U_jV_j^R.
\end{equation}
where $U_j$ and $V_j$ are $2\times2$ special unitary matrices and $V_j^R$ is
the dual of $V_j$, as defined in Eq.~(\ref{eq:dual}).

\begin{widetext}

For this parameterization, the normalized invariant measure of $S_j$ is given
by
\begin{equation}
\mathrm{d}\mu_{\beta}(S_j) =
p_{\beta}(\tau_j)\, 
\mathrm{d}\tau_j\, 
\frac{\mathrm{d}\phi_j}{2\pi}\,
\frac{\mathrm{d}\psi_j}{2\pi} 
\times \left\{
\begin{array}{lr}
\vspace{0.1cm}
1\, , & \quad \beta=1,\\
\vspace{0.1cm}
\displaystyle{ \frac{\mathrm{d}\phi'_j}{2\pi}\, 
\frac{\mathrm{d}\psi'_j}{2\pi},} & \beta=2\, ,\\
\displaystyle{\mathrm{d}\mu(U_j)\, \mathrm{d}\mu(V_j)}\, , & \beta=4,
\end{array}
\right.
\label{IM}
\end{equation}
where
\begin{equation}
p_{\beta}(\tau_j) = \frac{\beta}{2} \tau_j^{\beta/2-1},
\label{eq:ptauj}
\end{equation}
$\mathrm{d}\mu(U_j)$ and $\mathrm{d}\mu(V_j)$ are the invariant measures
of $U_j$ and $V_j$, respectively. As we will see in what follows, $f$ does not 
depend neither on $U_j$ nor $V_j$ such that we do not need to know explicitly 
the expressions for $\mathrm{d}\mu(U_j)$ and $\mathrm{d}\mu(V_j)$ .

In terms of the parameterization of Eq.~(\ref{SjP}) the voltage drop deviation, given in Eq.~(\ref{eq:f3}),
can be written as
\begin{equation}
\label{eq:frt}
f = 1 + 
\frac{\tau_1\tau_2-2\tau_2-2\tau_2\sqrt{1-\tau_1}\cos{(\psi_1+\psi'_1})}
{\tau_1+\tau_2-\tau_1\tau_2-\tau_1\sqrt{1-\tau_2}\cos{(\phi_2+\phi'_2)}
+\tau_2\sqrt{1-\tau_1}\cos{(\psi_1+\psi'_1})} . 
\end{equation}
\end{widetext}
Since $f$ does not depend neither on $U_j$ nor $V_j$, the integration over
$\mathrm{d}\mu(U_j)$ and $\mathrm{d}\mu(V_j)$ in Eq.~(\ref{eq:Pf1}) gives just
1 for $\beta=4$; except for variables $\tau_1$ and $\tau_2$, the resulting
expression for $P_{\beta}(f)$ reduces to a one similar to that for
$\beta=1$ in the remaining variables. The integration of
$\mathrm{d}\phi_1/2\pi$, $\mathrm{d}\phi'_1/2\pi$, $\mathrm{d}\psi_2/2\pi$, and
$\mathrm{d}\psi'_2/2\pi$ gives 1 for $\beta=2$. Therefore, once we perform the
integration with respect to $\phi_2$ and $\phi_2'$, making the appropriate change of
variables for each symmetry class, and then with respect to the remaining phases
$\psi_1$ and $\psi_1'$, we arrive at
\begin{equation}
P_{\beta}(f) = 
\int_0^1 \mathrm{d}\tau_1 \int_0^1 \mathrm{d}\tau_2 \,
p_{\beta}(\tau_1)\, p_{\beta}(\tau_2)\, p(f|\tau_1,\tau_2) ,
\label{Pf_betappp}
\end{equation}
where $p(f|\tau_1,\tau_2)$, which can be interpreted as the conditional probability 
distribution of $f$ given $\tau_1$ and $\tau_2$, is given by
\begin{eqnarray}\
\label{eq:p-conditional1}
p(f|\tau_1,\tau_1) &=& \frac{1}{\pi^2\left(1-f^2\right)
\sqrt{1-\tau_1}} \times \nonumber \\
&& \int_{L} 
\frac{2-\tau_1 + 2\sqrt{1-\tau_1}\,x}
{\sqrt{\left(\alpha_{+}-x\right)
\left(x-\alpha_{-}\right)
\left(1-x^2\right)}} \,
\mathrm{d}x , \nonumber \\
\end{eqnarray}
where $L\in(-1,1)\cap(\alpha_{-},\alpha_{+})$ with 
\begin{equation}
\alpha_{\pm} = 
\frac{(\tau_1-\tau_2)-(\tau_1 + \tau_2-\tau_1\tau_2)f\pm 
(1-f)\tau_1\sqrt{1-\tau_2}}
{(1+f)\tau_2\sqrt{1-\tau_1}},
\end{equation}
The integration in the domain $L$ is not a trivial problem because it depends on
the variables $\tau_1$, $\tau_2$, and $f$ through $\alpha_{\pm}$ in a
complicated way, but we can give a further step in the integration by looking
for the limits of integration once the value of $f$ is fixed (see Appendix~\ref{sec:AppendixA} for details). The integral over $x$ in Eq.~(\ref{eq:p-conditional1})
can be expressed in terms of complete elliptic integrals. The result for
$P_\beta(f)$ can be expressed as 
\begin{eqnarray}
P_{\beta}(f) &=& \frac{1}{\pi^2(1-f^2)} \times \nonumber \\
&& \left[ \int_0^1 \mathrm{d}\tau_1 
\frac{p_{\beta}(\tau_1)}{\sqrt{1-\tau_1}} 
\int_0^{u(\tau_1)} \mathrm{d}\tau_2 \,
p_{\beta}(\tau_2)\, I_{-1}^1(f;\tau_1,\tau_2) \right.
\nonumber \\ 
&+& \int_0^1 \mathrm{d}\tau_2\, p_{\beta}(\tau_2)\,
\int_0^{w(\tau_2)} \mathrm{d}\tau_1 
\frac{p_{\beta}(\tau_1)}{\sqrt{1-\tau_1}}\,
I_{\alpha_{-}}^{\alpha_{+}}(f;\tau_1,\tau_2) \nonumber
\label{eq:P(f)Elliptical}
\\ 
&+& H(f)\, H(1-f) 
\int_0^1 \mathrm{d}\tau_1 \frac{p_{\beta}(\tau_1)}{\sqrt{1-\tau_1}} \times \\
&& \int_{u(\tau_1)}^{v(\tau_1)}\, \mathrm{d}\tau_2\, 
p_{\beta}(\tau_2)\, I_{-1}^{\alpha_{+}}(f;\tau_1,\tau_2) 
\nonumber \\
&+& \left.
H(-f)\, H(f+1)
\int_0^1 \mathrm{d}\tau_1 
\frac{p_{\beta}(\tau_1)}{\sqrt{1-\tau_1}} \right. \times \nonumber \\
&& \left. \int_{u(\tau_1)}^{v(\tau_1)} \mathrm{d}\tau_2\, 
p_{\beta}(\tau_2)\, I_{\alpha_{-}}^1(f;\tau_1,\tau_2) 
\right], \nonumber
\end{eqnarray}
where $H(x)$ is the Heaviside step function and the limits are 
\begin{eqnarray}
\label{eq:limits}
u(\tau_1) &=& \frac{r(2s_{-}-r)}{s_{-}^2}, \quad 
v(\tau_1) = \frac{r(2s_{+}-r)}{s_{+}^2}, \quad \nonumber \\ 
&& \mathrm{and} \quad w(\tau_2) = \frac{r'(2s'-r')}{{s'}^2},
\end{eqnarray}
where $r'=r(\tau_1\to\tau_2)$ and $s'=s_{-}(\tau_1\to\tau_2)$ with
\begin{eqnarray}
r(\tau_1) &=& (1-f)\tau_1 \quad\mbox{and}\quad \nonumber \\
s_{\pm}(\tau_1) &=& 1 + (1-\tau_1)f \pm (1+f)\sqrt{1-\tau_1}.
\end{eqnarray}
In Eq.~(\ref{eq:P(f)Elliptical}), 
\begin{eqnarray}
& & I_{-1}^1(f;\tau_1,\tau_2) = 
\frac{2}{\sqrt{(\alpha_{+}+1)(1-\alpha_{-})}}  \nonumber \\ \times
&& [(A + B\alpha_{-})K(k_1)-(1+\alpha_{-})B\,\Pi(r_1^2,k_1)],  \nonumber \\
\label{C1}
\end{eqnarray}
\begin{eqnarray}
& & I_{\alpha_{-}}^{\alpha_{+}}(f;\tau_1,\tau_2) = 
\frac{2}{\sqrt{(\alpha_{+}+1)(1-\alpha_{-})}} \nonumber \\ \times
&& [(A - B)K(k_1)+(1+\alpha_{-})B\,\Pi({r'_1}^2,k_1)], \nonumber \\
\label{C2}
\end{eqnarray}
\begin{eqnarray}
& & I_{-1}^{\alpha_{+}}(f;\tau_1,\tau_2) =
\frac{\sqrt{2}}{\sqrt{\alpha_{+}-\alpha_{-}}} \nonumber \\ \times
&& [(A+ B \alpha_{-})K(k_2)-(1+\alpha_{-})B\,\Pi({r'_2}^2,k_2)], \nonumber \\
\label{C3}
\end{eqnarray}
and
\begin{eqnarray}
& & I_{\alpha_{-}}^1(f;\tau_1,\tau_2) =
\frac{\sqrt{2}}{\sqrt{\alpha_{+}-\alpha_{-}}} \nonumber \\ \times
&& [(A - B)K(k_2)+(1+\alpha_{-})B\,\Pi({r_2}^2,k_2)], \nonumber \\
\label{C4}
\end{eqnarray}
where $A=2-\tau_1$ and $B=2\sqrt{1-\tau_1}$; $K(k)$ and $\Pi(r^2,k)$ are 
complete elliptic integrals of the first and third kind, respectively, and 
\begin{eqnarray}
k_1^2 &=& \frac{2(\alpha_{+}-\alpha_{-})}{(\alpha_{+}+1)(1-\alpha_{-})}, \quad 
r_1^2 = \frac{2}{1-\alpha_{-}}, \nonumber \\ && \quad\mbox{and}\quad 
{r'_1}^2 = \frac{\alpha_{+}-\alpha_{-}}{\alpha_{+}+1}; 
\end{eqnarray}
\begin{equation}
k_2^2 = \frac{1}{k_1^2}, \quad r_2^2 = \frac{1}{r_1^2}, 
\quad\mbox{and}\quad {r'_2}^2 = \frac{1}{{r'_1}^2}.
\end{equation}

\begin{figure}
\centering
\includegraphics[width=1.0\columnwidth]{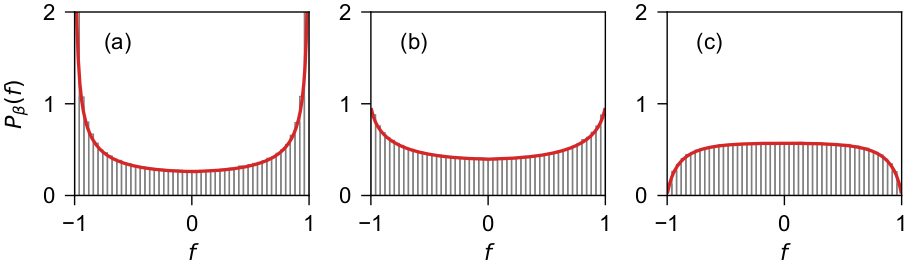}
\caption{Comparison between the analytical results
(continuous lines) with random matrix simulations (histograms) for
$P_{\beta}(f)$: (a) $\beta=1$, (b) $\beta=2$, and (c) $\beta=4$.}
\label{fig:Pfbeta}
\end{figure}

The remaining integrals with respect to $\tau_1$ and $\tau_2$ can be performed
numerically. In order to verify our results, in Fig.~\ref{fig:Pfbeta} we
compare them with the statistical distributions obtained from numerical
simulations using Eqs.~(\ref{SjP})-(\ref{eq:frt}). An excellent agreement is
observed.

We can observe in Fig.~\ref{fig:Pfbeta} that the distribution of $f$ is
symmetric with respect to $f=0$. Moreover, it diverges at $f=\pm1$ for 
$\beta=1$, while it shows finite peaks at $f=\pm1$ for $\beta=2$, which is a 
clear effect of the weak localization. For $\beta=4$ the distribution becomes 
zero at $f=\pm1$ due to the antilocalization phenomenon. The differences found 
in the $P_{\beta}(f)$ between the symmetry classes are important
signatures of the chaotic setup we consider here, since in the equivalent 
disordered configuration no differences were found~\cite{Godoy1992b}.


\section{Experimental realizations with microwave graphs}
\label{sec:exp}

\begin{widetext}

\begin{figure}
\centering
\includegraphics[width=1.0\columnwidth]{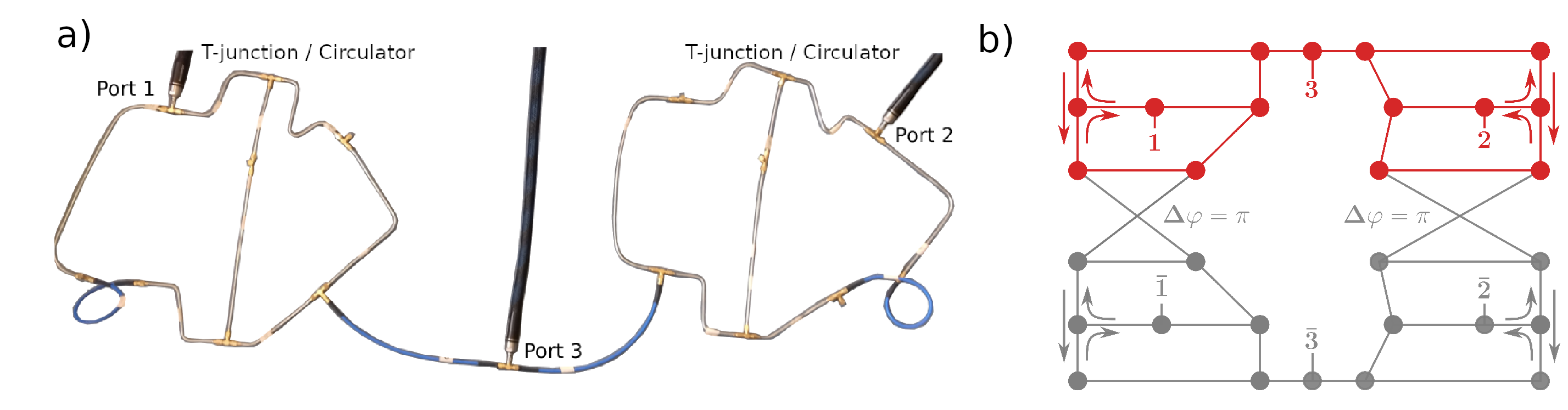}
\caption{(a) Photograph of the three-port experimental setup  for 
the case of time-reversal symmetry ($\beta=1$). For the realization of a break 
of time-reversal symmetry ($\beta=2$) one T junction in each of the subgraphs is 
replaced by a circulator. Both subgraphs are geometrically different but have 
the same  total length, i.\,e.,  the  mean level density is the same. (b) Sketch 
of the graph for $\beta=1$ (GOE)  and $\beta=2$ (GUE), where for $\beta=1$ the 
circulators are replaced by ordinary T junctions (red). For a realization of 
$\beta=4$ (GSE) each GUE subgraph is complemented by another geometrically 
identical one but with an opposite sense of rotation of the circulators (grey). 
The two respective copies are connected by pairs of bonds with length 
differences corresponding to a phase difference of $\pi$ for the propagating 
waves (see Ref.~\cite{Abdu2016} for details, where also a photograph of a single 
GSE graph can be found).}
\label{fig:Exp}
\end{figure}

\end{widetext}

Since the voltage drop deviation depends only on the scattering properties of 
the devices, it can be emulated in classical wave systems, in particular 
experiments with microwave graphs can be performed. 
Figure~\ref{fig:Exp}(a) shows a photograph of the experimental set-up for the 
case of time-reversal invariance ($\beta=1$). The devices consist of chaotic 
microwave networks formed by coaxial semirigid cables (Huber \& Suhner
EZ-141) with SMA connectors, coupled by T junctions at the nodes. 
One microwave port attached to the left subgraph acts as the 
input, another one on the right as the output, and a third port attached to the 
connecting cable between the two subgraphs as the probe. To realize a break of 
time-reversal invariance ($\beta=2$), in each of the two subgraphs one  of the T 
junctions is replaced by a circulator (Aerotek I70-1FFF) with an operating 
frequency range from 6 to 12 GHz. A circulator introduces directionality, waves 
entering via port 1 leave via port 2. The transmission intensities $T_{31}$ and 
$T_{32}$ were measured by an Agilent 8720ES vector network analyzer (VNA). 
The $\beta=1$ case has been additionally realized  in a billiard 
setup. Details are presented in Appendix~\ref{sec:AppendixB}.

For the realization of the $\beta = 4$ (GSE) case two GSE graphs are needed. 
Since each GSE graph is composed by two GUE subgraphs, representing the spin-up 
and spin-down components \cite{Abdu2016}, we need a total of four subgraphs. 
Figure~\ref{fig:Exp}(b) shows a sketch of the three-terminal GSE graph used to 
measure the scattering matrix. An essential ingredient of the setup are two 
pairs of bonds with length differences $\Delta l$ corresponding to phase 
differences of $\Delta\varphi=\pi$ for the propagating waves. In the experiment 
we took spectra for fixed $\Delta l$  and converted them into spectra for fixed 
$\Delta\varphi$ using $\Delta\varphi=k\Delta l$, where $k$ is the wavenumber.  
Details can be found in Ref.~\cite{Abdu2016}. The ports now appear in pairs 
$1, \bar{1}, 2, \bar{2}, 3, \bar{3}$, and the scattering matrix elements turn 
into $2\times 2$ matrices
\begin{eqnarray}
\mathbf{S}_{ij} =
\left(
\begin{array}{cc}
S_{ij} & S_{i\bar{j}} \\
S_{\bar{i}j} & S_{\bar{i}\bar{j}}
\end{array}
\right) .
\end{eqnarray}
In the spirit of the spin analogy, $S_{ij}$ and $S_{i\bar{j}}$ correspond to 
transmissions without and with spin-flip, respectively. $\mathbf S_{ij}$ may be 
written in terms of quaternions as 
$\mathbf{S}_{ij}=S^0_{ij}\openone+\mathrm{i}S^x_{ij}\sigma_x+\mathrm{i}S^y_{ij}
\sigma_y+\mathrm{i}S^z_{ij}\sigma_z$, where for a symplectic symmetry all 
coefficients $S^0_{ij}$, $S^x_{ij}$, etc. are real numbers and $\sigma_j$ is 
the corresponding Pali matrix $j$. As a consequence $\mathbf S_{ij}\mathbf 
S_{ij}^\dag$ is a multiple of the unit matrix which allows for a simple check of 
the quality of the realization of the setup with symplectic symmetry. In our 
experiments $\mathbf{S}_{31}$ and $\mathbf{S}_{32}$ were found to be quaternion 
real within an error of $2.19\%$ and $1.93\%$, respectively. The mentioned 
multiple is nothing but the transmission coefficient, hence 
$T_{31}=\frac{1}{2}\mathrm{tr}(\mathbf{S}_{31}\mathbf{S}_{31}^{\dagger})$ and 
$T_{32}=\frac{1}{2}\mathrm{tr}(\mathbf{S}_{32}\mathbf{S}_{32}^{\dagger})$.

\begin{figure}
\centering
\includegraphics[width=1\columnwidth]{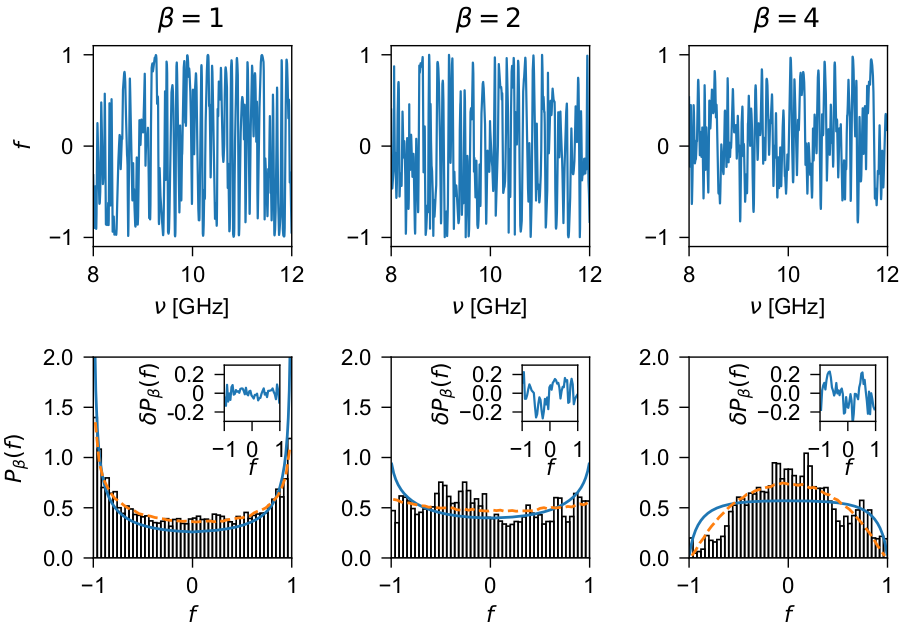}
\caption{(Color online) Experimental voltage drop deviation $f$ as a function of frequency is shown in
the upper panels for the three symmetry classes. Their corresponding
statistical distribution are shown as histograms. The continuous (blue) lines
in the lower panels correspond to the theoretical result shown in
Fig~\ref{fig:Pfbeta}, while the dashed (orange) lines are the RMT simulations
which take into account the dissipation and imperfect coupling: $\beta=1$
(left), $\beta=2$ (middle), and $\beta=4$ (right). For comparison purposes, in the insets in
the lower panels we show the difference between the numerical and the experimental
distribution $\delta P_{\beta}(f) = P_{\beta}(f)_{\mathrm{num}} - P_{\beta}(f)_{\mathrm{expt}}$.
For the statistical analysis we used an ensemble of $5 \times 10^{4}$ realizations.}
\label{fig:Pb(f)Exp}
\end{figure}

The experimental distribution of $f$ is shown in Fig.~\ref{fig:Pb(f)Exp} as
histograms for the three symmetry classes $\beta$, where it is compared with the
theoretical result given by Eq.~(\ref{eq:P(f)Elliptical}). As can be observed
there is a qualitatively good agreement for all $\beta$. Despite the fact that there is a
quantitative difference between theory and experiment, which is due to the phenomena of
dissipation and imperfect coupling between the graphs and ports, this difference is not so large. In
Fig.~\ref{fig:Pb(f)Exp} we also show the corrected distribution obtained from
RMT simulations, using the Heidelberg approach (see below), once these two
phenomena, dissipation and imperfect coupling, are taken into account.

\begin{figure}
\centering
\includegraphics[width=1.0\columnwidth]{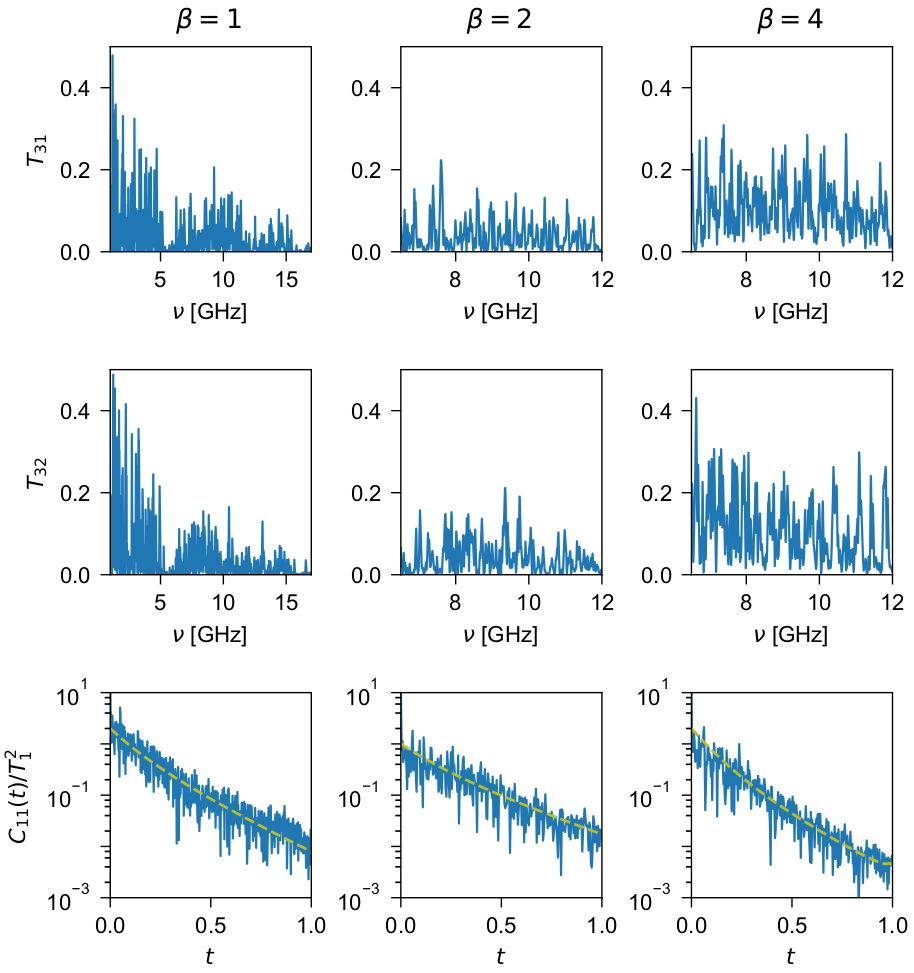}
\caption{Experimental transmission intensities $T_{31}$ and $T_{32}$ as a
function of frequency are shown in upper and middle panels for the three
symmetry classes. The lower panels show the autocorrelation function: the
fluctuations are the experimental measurements and the dashed lines are the best fits of
Eq.~(\ref{eq:corr}) to the data for $t<1$.}
\label{fig:Tij}
\end{figure}

The dissipation and imperfect coupling can be quantified by two parameters: $T_1$ for the coupling
strength and $\gamma$ for the dissipation. The coupling between the graphs and
ports $T_1$ is extracted from the experimental data as $T_1=1-|\langle
S_{11}\rangle|^2$, where $\langle S_{11}\rangle$ is the average (with respect to the frequency) of
the scattering matrix element 11. Here we obtained
$T_1=0.99$ for $\beta=1$, $T_1=0.99$ for $\beta=2$, and $T_1=0.953$ for
$\beta=4$. As can be seen the coupling strength is almost 1, which means that
the coupling between graphs and ports, although not perfect, can be considered as very good.
The effect of dissipation is clearly observed in the experimental
transmission intensities $T_{31}$ and $T_{32}$. Their fluctuations as a
function of the frequency are shown in Fig.~\ref{fig:Tij}, where we observe
that they do not reach the value 1. The dissipation parameter $\gamma$ can be quantified by fitting the autocorrelation
function $C_{11}(t)$ of the 11 element of the scattering matrix~\cite{Schafer2003,Angel2018,Fyodorov2005b}
\begin{equation}
\label{eq:corr}
\frac{C_{11}(t)}{T_1^2} =
\left\{
\begin{array}{ll}
\vspace{0.1cm}
\left[ \displaystyle{ \frac{3}{(1+2T_1t)^3}-
\frac{b_{1,2}(t)}{(1+T_1t)^4}} \right] 
\mathrm{e}^{-\gamma t}
 & \mbox{for} \, \, \beta=1, \\ \vspace{0.1cm}
\left[ \displaystyle{ \frac{2}{(1+T_1t)^4}-
\frac{2^6b_{2,2}(t)}{(2+T_1t)^6}} \right] 
\mathrm{e}^{-\gamma t}
 & \mbox{for} \, \, \beta=2, \\
 \left[ \displaystyle{ \frac{6}{(1+T_1t)^6}-
\frac{2^{12}b_{4,2}(t)}{(2+T_1t)^{10}}} \right] 
\mathrm{e}^{-2\gamma t}
 & \mbox{for} \, \, \beta=4,
\end{array}
\right.
\end{equation}
where $b_{\beta,2}(t)$ is the two-level form factor~\cite{Guhr1998} and $T_{1}$, $\gamma$, and $t$ are given in dimensionless units. The
best fit achieved for $C_{11}(t)$, for $t<1$, is shown in Fig.~\ref{fig:Tij} for all
symmetry classes, for which we obtain $\gamma=2.6$ for $\beta=1$,
$\gamma=2.0$ for $\beta=2$, and $\gamma=1.8$ for $\beta=4$.

Once these parameters are determined, RMT simulations can be performed for the
scattering matrix of each graph, assumed as quantum systems. In the Heidelberg
approach the scattering matrix of a graph can be generated as~\cite{Brouwer1997}
\begin{equation}
S_g = 1 - 2\pi\mathrm{i} W^{\dagger}
\frac{1}{E-\mathcal{H}+\mathrm{i}\pi WW^{\dagger}} W,
\end{equation}
where $E$ represents the energy of the incoming wave and $W$ is the matrix that couples the open modes
in the ports to the internal modes of the system. $\mathcal{H}$ is an effective Hamiltonian
that includes the dissipation $\gamma$,
$\mathcal{H_{\mu\nu}}=H_{\mu\nu}-\mathrm{i}\delta_{\mu\nu}\gamma\Delta/4\pi$, with $\Delta$ the mean level spacing.
The imperfect coupling can be modeled by adding identical barriers, with transmission intensity $T_{1}$, between the graph 1 and port 1, between the graphs and the T junction, and between graph 2 and port 2.
The numerical simulations of the three-terminal system, with the obtained values of $T_1$ and
$\gamma$, lead us to the result shown in Fig.~\ref{fig:Pb(f)Exp} for the
distribution of $f$; see the orange dashed lines in the lower panels. As can be observed, the agreement is very good.


\section{Conclusions}
\label{sec:conclusions}

A three-terminal symmetrical system consisting of two microwave graphs was used to measure
indirectly the voltage drop between the graphs. One port was used as an
input, a second port as an exit, and a third port as a probe. The fluctuations
of the quantity $f$, that accounts for the deviation from the mean value of the
potentials, are explained qualitatively by an ideal result obtained from the
scattering approach of random matrix theory for the three Dyson's symmetry classes. We
found that the distribution of $f$ is symmetric with respect to zero, whose shape is
quite similar to the one obtained in the disordered case, in the insulating regime, but with an
important difference which is the effect of weak localization and
antilocalization phenomena not found in the disordered case. A more accurate
description is obtained when dissipation and imperfect coupling between the ports
and graphs are taken into account.


\acknowledgments

This work was supported by CONACyT (Grant No. CB-2016/285776).
F.~C.-R. thanks financial support from CONACyT and A.M.M-A. acknowledges
support from DGAPA-UNAM. J.A.M.-B. acknowledges financial support from CONACyT (Grant No. A1-S-22706). The experiments were funded by the Deutsche Forschungsgemeinschaft via the individual grants STO 157/17-1 and KU 1525/3-1 including a short-term visit of A.M.M.-A. in Marburg.


\appendix
\addcontentsline{toc}{chapter}{Apendices}

\section{Limits of integration in Eq.~(\ref{eq:p-conditional1})}
\label{sec:AppendixA}

The integration with respect to $x\in(-1,1)$ in Eq.~(\ref{eq:p-conditional1})
should be performed in the interval $L$, which is the intersection of the
intervals $(-1,1)$ and $(\alpha_{-},\alpha_{+})$. Since $\alpha_{+}$ and
$\alpha_{-}$ depend on the values of $\tau_1$, $\tau_2$, and $f$, the values of
$\tau_1$ and $\tau_2$ may be affected by that intersection for fixed $f$. This
leads to integrals of the form
\begin{equation}
I_{a}^{b}(f,\tau_1,\tau_2)= \int_{a}^{b} \frac{(A + B 
x)dx}{\sqrt{(\alpha_{+}-x)(x-\alpha_{-})(1+x)(1-x)}},
\label{Ix}
\end{equation}
which gives rise to complete elliptic integrals of the first and third
kind.

\begin{figure}
\centering
\includegraphics[width=0.8\columnwidth]{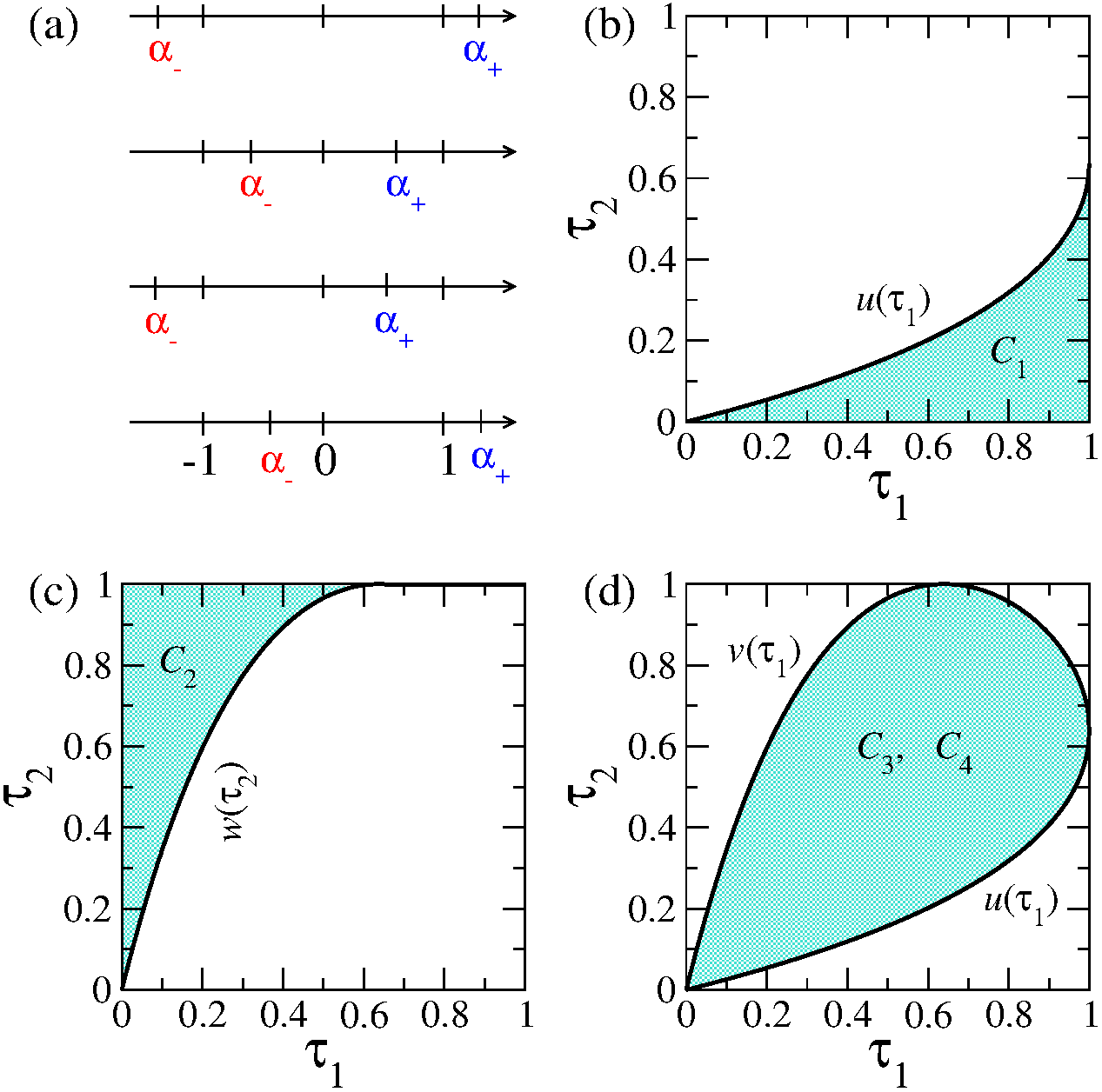}
\caption{(a) Intervals of integration for $x$ which imply restrictions on
$\tau_1$ and $\tau_2$ (shaded regions): for $|f|=0.6$ (b)
$\tau_2\in(0,u(\tau_1))$ with $\tau_1\in(0,1)$; (c) $\tau_1\in(0,w(\tau_2))$
with $\tau_2\in(0,1)$; (d) $\tau_2\in(u(\tau_1), v(\tau_1))$ for $f>0$ and
$\tau_2\in(v(\tau_1),u(\tau_1))$ for $f<0$, with $\tau_1\in(0,1)$. The functions
$u(\tau)$, $v(\tau)$, and $w(\tau)$ are defined in Eq.~(\ref{eq:limits}).}
\label{fig:Intervalos}
\end{figure}

Four conditions arise, as illustrated in Fig.~\ref{fig:Intervalos}(a). Restrictions
on $\tau_1$ and $\tau_2$ are obtained for a fixed value of $f$. These
restrictions lead to the several regions in the plane $\tau_1\tau_2$, as can be
seen in Fig.~\ref{fig:Intervalos}.

\paragraph{Condition $C_1$} $\alpha_{-}\leq-1$ and $1\leq\alpha_{+}$. Under
these conditions $x$ runs over its full domain, $x\in(-1,1)$. Restriction on
$\tau_2$ for a fixed value of $f$ is obtained from these two conditions.
The first condition, $\alpha_{-}\leq-1$, leads to $\tau_2\ge 0$ while the
second condition, $1\leq\alpha_{+}$, restricts $\tau_2$ to be smaller than
$u(\tau_1)$, where $u(\tau_1)$ is given in equation of (\ref{eq:limits}).
Therefore, $\tau_2\in(0,u(\tau_1))$ with $\tau_1\in(0,1)$.

\paragraph{Condition $C_2$} $-1\leq\alpha_{-}\leq\alpha_{+}\leq 1$. For this
condition $x\in(\alpha_{-},\alpha_{+})$. Here, it is more 
convenient to see the restriction on $\tau_1$ for which we obtain that 
$\tau_1\in(0,w(\tau_2))$, where $w(\tau_2)$ is given by the third equation in 
(\ref{eq:limits}), for $\tau_2\in(0,1)$.

\paragraph{Condition $C_3$} $\alpha_{-}\leq-1$ and $-1\leq\alpha_{+}\leq 1$.
The integration over $x$ is defined in the interval $(-1,\alpha_{+})$. The
first part of the condition is the same as that for Condition $C_1$, which
restricts to $\tau_2\ge 0$. The second condition has two parts, a first one is
$\alpha_{+}\leq 1$ whose result is similar to the corresponding second part in
Condition $C_1$, but the inequality in opposite sense; hence, $\tau_2\geq
u(\tau_1)$. In a similar manner, the second part $\alpha_{+}\ge-1$ leads to
$\tau_2\leq v(\tau_1)$ for $f>0$, which is given by the second equation in
(\ref{eq:limits}). Therefore, $\tau_2\in(u(\tau_1), v(\tau_1))$ with
$\tau_1\in(0,1)$ and $f>0$.

\paragraph{Condition $C_4$} $-1\leq\alpha_{-}\leq1$ and $1\leq\alpha_{+}$.
The integration over $x$ is defined in the interval $(\alpha_{-},1)$. The
second part of the condition is the same as that for Condition $C_1$ which
restricts to $\tau_2\in(0,u(\tau_1))$ if $f$ is positive. The first condition
has two parts, the first one, $\alpha_{-}\leq 1$ leads to the same restriction
to $\tau_2$. The second part gives $\tau_2\in(v(\tau_1),1)$ for $f>0$. This
result is equivalent to $\tau_2\in(u(\tau_1),v(\tau_1))$ for $f<0$.\\

To apply these conditions to the integral of Eq.~(\ref{Ix}) we use the
equations 254.00, 254.10, 336.01, 336.60, and 340.04 of Ref.~\cite{ByrdHandbook},
then we integrate with respect to $\tau_1$ and $\tau_2$, and finally we arrive at
Eq.~(\ref{eq:P(f)Elliptical}).


\section{Three-port microwave billiard experiment}
\label{sec:AppendixB}

\begin{figure}
\centering
\includegraphics[width=1.0\columnwidth]{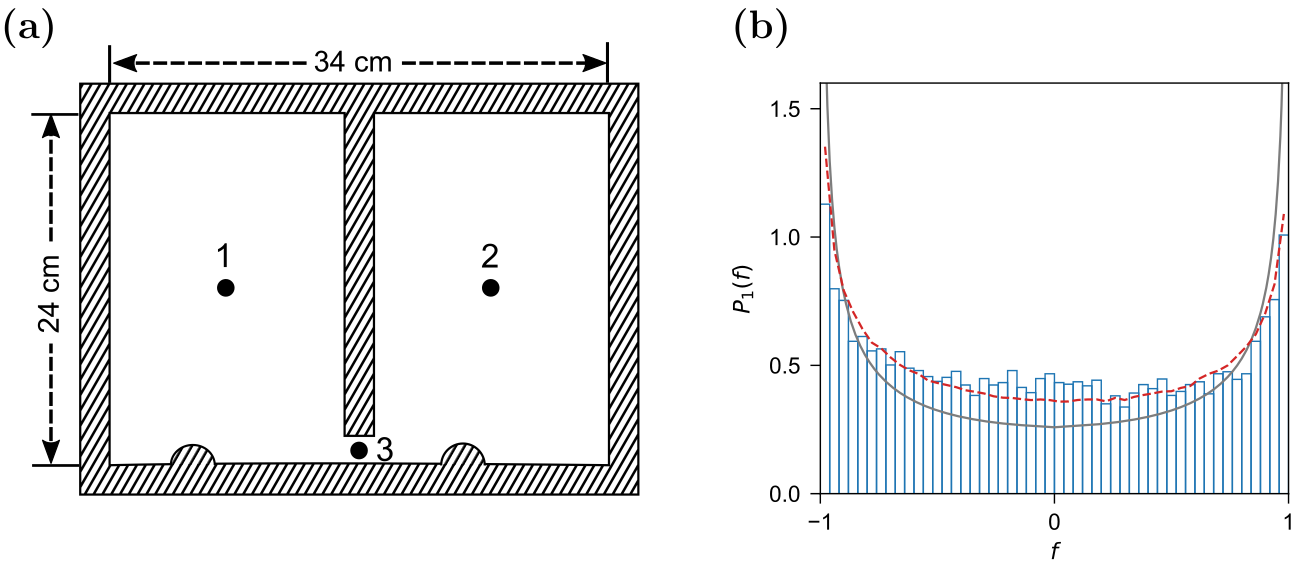}
\caption{(a) Sketch of the three-port microwave billiard in the 
presence of $\beta=1$ symmetry. The ports are labeled as 1, 2, and 3, where port 
3 is used as a probe. The experimental transmission intensities $T_{31}$ and 
$T_{32}$ are measured between ports 1 and 3, and between ports 2 and 3, 
respectively. (b) Distribution $P_{1}(f)$: the continuous (black) line 
corresponds to the theoretical result shown in Fig.~\ref{fig:Pfbeta}, the dashed 
(red) line is the RMT simulation which takes into account 
dissipation and imperfect coupling, while the histogram corresponds to the 
experimental result.}
\label{fig:3TBilliard}
\end{figure}

For the $\beta=1$ case, in addition to the microwave graph experiments an 
experiment in a billiard setup has been performed. A sketch is shown in 
Fig.~\ref{fig:3TBilliard}(a). The billiard is constructed on an aluminum plate 
with two sub-billiards of the same shape separated by a central bar.  The mirror 
symmetry is broken by two semicircular obstacles, attached to the bottom 
boundary. The ports are labeled as 1, 2, and 3, where port 3 is used as a probe. 
The experimental transmission intensities $T_{31}$ and $T_{32}$ have been 
measured between port 1 and port 3, and between port 2 and port 3, respectively, 
for frequencies from 1 to 17 GHz. With a distance of $d=8$\,mm between top and 
bottom plate the billiard is quasi-two-dimensional in the whole frequency 
range.

In panel (b) of Fig.~\ref{fig:3TBilliard} we show as histograms the experimental 
distribution of $P_{1}(f)$ obtained from the graph (black) and the billiard 
(blue) setting. The analytical result is shown in the continuous (black) line, 
while the RMT simulation, which take into account the dissipation and imperfect 
coupling, is shown in (red) dashed line. Again, a good agreement between 
experiment and theory is found for $P_{1}(f)$.




\begin{thebibliography}{99}

\bibitem{Fyodorov2005a}
Y. V. Fyodorov, T. Kottos, H.-J. St\"ockmann,
Trends in quantum chaotic scattering,
J. Phys. A: Math. Gen. \textbf{38} (2005) Preface.

\bibitem{Schanze2001}
H. Schanze, E. R. P. Alves, C. H. Lewenkopf, and H.-J. St\"ockmann,
Phys. Rev. E \textbf{64}, 065201(R) (2001).

\bibitem{Schanze2005}
H. Schanze, H.-J. St\"ockmann, M. Mart\'inez-Mares, and C. H. Lewenkopf,
Phys. Rev. E \textbf{71}, 016223 (2005).

\bibitem{Schafer2003}
R. Sch\"afer, T. Gorin, T. H. Seligman, and H.-J. St\"ockmann,
J. Phys. A: Math. Gen. \textbf{36}, 3289 (2003).

\bibitem{Kuhl2005}
U. Kuhl, M. Mart\'inez-Mares, R. A. M\'endez-S\'anchez, and H.-J. St\"ockmann,
Phys. Rev. Lett. \textbf{94}, 144101 (2005).

\bibitem{Buttiker1988}
M. B\"uttiker,
IBM J. Res. Develop. \textbf{32}, 317 (1988).

\bibitem{Schwab2000}
K. Schwab, E. A. H. J. M. Worlock, and M. L. Roukes,
Nature \textbf{404}, 974 (2000).

\bibitem{Buttiker1984}
M. B\"uttiker, Y. Imry, and M. Y. Azbel,
Phys. Rev. A \textbf{30}, 1982 (1984).

\bibitem{Brouwer1997}
P. W. Brouwer and C. W. J. Beenakker,
Phys. Rev. B \textbf{55}, 4695 (1997).

\bibitem{Keller1996}
M. W. Keller, A. Mittal, J. W. Sleight, R. G. Wheeler, D. E. Prober, R. N. Sacks, and H. Shtrikmann,
Phys. Rev. B \textbf{53}, R1693 (1996).

\bibitem{Marcus1992}
C. M. Marcus, A. J. Rimberg, R. M. Westervelt, P. F. Hopkins, and A. C. Gossard,
Phys. Rev. Lett. \textbf{69}, 506 (1992) .

\bibitem{Chan1995}
I. H. Chan, R. M. Clarke, C. M. Marcus, K. Campman, and A. C. Gossard,
Phys. Rev. Lett. \textbf{74}, 3876 (1995).

\bibitem{Moises2005}
M. Mart\'inez-Mares,
Phys. Rev. E \textbf{72}, 036202 (2005).

\bibitem{Enrique2016}
E. Flores-Olmedo, A. M. Mart\'inez-Arg\"uello, M. Mart\'inez-Mares, G. B\'aez, J. A. Franco-Villafa\~ne, and R. A. M\'endez-S\'anchez,
Sci. Rep. \textbf{6}, 25157 (2016).

\bibitem{Buttiker1986}
M. B\"uttiker,
Four-terminal phase-coherent conductance,
Phys. Rev. Lett. \textbf{57}, 1761 (1986).

\bibitem{Gopar1994}
V. A. Gopar, M. Mart\'inez, and P. A. Mello,
Phys. Rev. B \textbf{50}, 2502 (1994).

\bibitem{Arrachea2008}
L. Arrachea, C. Na\'on, and M. Salvay,
Phys. Rev. B \textbf{77}, 233105 (2008).

\bibitem{Texier2016}
C. Texier and G. Montambaux,
Physica E \textbf{82}, 272 (2016).

\bibitem{Foieri2009}
F. Foieri, L. Arrachea, and M. J. S\'anchez,
Phys. Rev. B \textbf{79}, 085430 (2009).

\bibitem{DAmato1990}
J. L. D'Amato and H. M. Pastawski,
Phys. Rev. B \textbf{41}, 7411 (1990).

\bibitem{Cattena2014}
C. J. Cattena, L. J. Fern\'andez-Alc\'azar, R. A. Bustos-Mar\'un, D. Nozaki, and H. M. Pastawski,
J. Phys.: Condens. Matter \textbf{26}, 345304 (2014).

\bibitem{Song1998}
A. M. Song, A. Lorke, A. Kriele, J. P. Kotthaus, W. Wegscheider, and M. Bichler,
Phys. Rev. Lett. \textbf{80}, 3831 (1998).

\bibitem{Gao2005}
B. Gao, Y. F. Chen, M. S. Fuhrer, D. C. Glattli, and A. Bachtold,
Phys. Rev. Lett. \textbf{95}, 196802 (2005).

\bibitem{Angel2018}
A. M. Mart\'inez-Arg\"uello, A. Rehemanjiang, M. Mart\'inez-Mares, J. A. M\'endez-Berm\'udez, H.-J. St\"ockmann, and U. Kuhl,
Phys. Rev. B \textbf{98}, 075311 (2018).

\bibitem{Angel2019}
A. M. Mart\'inez-Arg\"uello, J. A. M\'endez-Berm\'udez, and M. Mart\'inez-Mares,
Phys. Rev. E \textbf{99}, 062202 (2019).

\bibitem{Godoy1992a}
S. Godoy and P. A. Mello,
EPL \textbf{17}, 243 (1992).

\bibitem{Godoy1992b}
S. Godoy and P. A. Mello,
Phys. Rev. B \textbf{46}, 2346 (1992).

\bibitem{Angel2014}
A. M. Mart\'inez-Arg\"uello, E. Casta\~no, and M. Mart\'inez-Mares,
Random matrix study for a three-terminal chaotic device, in \emph{Special Topics on Transport Theory: Electrons, Waves, and Diffusion in Confined Systems}, AIP Conf. Proc. No. 1579 (AIP, Melville, NY , 2014), p. 46.

\bibitem{RMP69}
C. W. J. Beenakker, 
Rev. Mod. Phys. \textbf{69}, 731 (1997).

\bibitem{Karol}
K. {\.Z}yczkowski, \emph{Random Matrices of Circular Symplectic Ensemble} in 
\emph{Chaos--The Interplay Between Stochastic and Deterministic Behaviour},  
Lecture Notes in Physics, vol 457, edited by P. Garbaczewski, M. Wolf, and  
A. Weron (Springer, Berlin, Heidelberg, 1995).

\bibitem{Abdu2018}
A. Rehemanjiang, M. Richter, U. Kuhl, and H.-J. St\"ockmann,
Phys. Rev. E \textbf{97}, 022204 (2018).

\bibitem{Abdu2016}
A. Rehemanjiang, M. Allgaier, C. H. Joyner, S. M\"uller, M. Sieber, U. Kuhl, and H.-J. St\"ockmann,
Phys. Rev. Lett. \textbf{117}, 064101 (2016).

\bibitem{Fyodorov2005b}
Y. V. Fyodorov, D. V. Savin, and H.-J. Sommers,
J. Phys. A: Math. Gen. \textbf{38}, 10731 (2005).

\bibitem{Guhr1998}
T. Guhr, A. M\"uller-Groeling, and H. A. Weidenm\"uller,
Phys. Rep. \textbf{299}, 189 (1998).

\bibitem{ByrdHandbook}
P. F. Byrd and M. D. Friedman,
Handbook of Elliptic Integrals for Engineers and Physicists, Springer-Verlag Berlin Heidelberg GMBH, Berlin, Heidelberg, 1954.

\end{thebibliography}
\end{document}